\documentclass[10pt,letterpaper]{article}
\usepackage{opex3}
\usepackage{cite}
\begin{document}
\title{Generation of broadband spontaneous parametric fluorescence 
using multiple bulk nonlinear crystals}

\author{Masayuki Okano,$^{1,2}$ Ryo Okamoto,$^{1,2}$ 
Akira Tanaka,$^{1,2}$ Shanthi Subashchandran,$^{1,2}$ and 
Shigeki Takeuchi$^{1,2,*}$}

\address{$^1$Research Institute for Electronic Science, Hokkaido University,
Sapporo 001-0020, Japan
 \\ $^2$The Institute of Scientific and Industrial Research, Osaka University,
Osaka 567-0047, Japan}

\email{$^*$takeuchi@es.hokudai.ac.jp} 


\begin{abstract*}
We propose a novel method for generating broadband spontaneous parametric fluorescence
by using a set of bulk nonlinear crystals (NLCs).
We also demonstrate this scheme experimentally. 
Our method employs a superposition of spontaneous parametric fluorescence spectra
generated using multiple bulk NLCs.
A typical bandwidth of 160 nm (73 THz) with a degenerate wavelength of 808 nm 
was achieved using two $\beta$-barium-borate (BBO) crystals, 
whereas a typical bandwidth of 75 nm (34 THz) was realized using a single BBO crystal.
We also observed coincidence counts of generated photon pairs 
in a non-collinear configuration.
The bandwidth could be further broadened by increasing the number of NLCs.
Our demonstration suggests that a set of four BBO crystals could realize 
a bandwidth of approximately 215 nm (100 THz).
We also discuss the stability of Hong-Ou-Mandel two-photon interference 
between the parametric fluorescence generated by this scheme.
Our simple scheme is easy to implement with conventional NLCs
and does not require special devices.\\
\end{abstract*}

\ocis{(190.4410) Nonlinear optics, Parametric processes; 
(270.0270) Quantum optics.} 



\section{Introduction}

Quantum entangled photon pairs play a key role
in quantum information processing \cite{knill2001scheme} 
and quantum communication \cite{duan2001long}.
In recent years, 
entangled photon pairs with ultrabroad bandwidths in the frequency domain
have attracted much attention 
\cite{{PhysRevLett.98.063602},{Brida2009}}.
Such frequency-entangled photon pairs are indispensable 
for the future realization of "single-cycle biphotons"
\cite{PhysRevLett.98.063602}, 
whose characteristic coincidence time, 
as measured at distant detectors, 
is a single optical cycle.
These broadband correlated photon pairs 
have the potential to realize
high-resolution quantum optical coherence tomography
\cite{PhysRevLett.91.083601},
highly-efficient two-photon absorption experiments 
\cite{{PhysRevLett.93.023005},{PhysRevLett.105.173602}}, and
precise synchronization of two clocks \cite{PhysRevLett.87.117902}.

Various methods have been developed 
for generating broadband spontaneous parametric fluorescence
that can be used as correlated photon pairs.
However, there have been only a few methods
that can realize an ultrabroad bandwidth of an order of 100 THz.
For example, one method uses chirped quasi-phase matched devices.
It was theoretically proposed by S.E. Harris \cite{PhysRevLett.98.063602}
and subsequently demonstrated experimentally by generating
ultrabroadband spontaneous parametric fluorescence 
in the range 700-1500 nm with a center wavelength $\lambda_c$ of 1064 nm 
\cite{mohan2009ultrabroadband}
and 300 nm (136 THz) bandwidth spontaneous parametric fluorescence
at $\lambda_c$ = 812 nm \cite{nasr2008ultrabroadband}.
Although this approach is suitable for ultrabroadband generation
and enables us to engineer the high optical nonlinearities of materials
\cite{branczyk2011engineering},
it requires sophisticated quasi-phase matched devices 
that should be properly designed for a specific purpose.
Furthermore, it is difficult to use this method to the applications
where the pump laser wavelength has to be varied 
over a certain range ($>$10 nm).
Another method employs a large temperature modulation in a NLC 
and obtained a broad bandwidth of 253 nm (154 THz) 
at $\lambda_c$ = 702 nm \cite{Katamadze2011TempControl}.
However, this method also requires sophisticated device structure
to exploit spatial modulation of temperature in a tiny NLC.
Although a broad bandwidth of 174 nm (106 THz) at $\lambda_c$ = 702 nm
was reported using an ultrathin (thickness: 0.05 mm) crystal \cite{dauler1999tests},
this method increases the bandwidth at the expense of the photon flux.

There have been other experimental demonstrations
towards the generation of broadband spontaneous parametric fluorescence.
One approach is to control the pump beam incident angles 
to obtain a wide range of the phase matching conditions.
A method using tightly focused pump beam 
was reported with 84 THz bandwidth \cite{Carrasco2006focus}, 
and another method uses a pair of gratings 
for pump (24 THz \cite{hendrych2009broadening}).
Oppositely, one can collect the parametric fluorescence
spread over in a certain divergence angle (91 THz \cite{o2007observation}).

In this paper, 
we propose a novel method that uses multiple bulk NLCs
to generate broadband spontaneous parametric fluorescence.
Our scheme is easy to implement as it employs a set of conventional bulk NLCs
and it does not require special devices.
This scheme uses a weakly focused pump beam 
and the broadband photons are emitted into a certain direction in a non-collinear configuration,
so that the parametric fluorescence can be coupled into single mode fiber couplers.  
In principle, the bandwidth can be extended by increasing the number of NLCs
up to 100 THz as is shown later.
Furthermore, the pump laser frequency can be tuned
within the tuning range of the phase-matching conditions of NLCs.
We then experimentally demonstrate the scheme by using two bulk NLCs.
A typical bandwidth of 160 nm (73 THz) with a degenerate wavelength of 808 nm
was achieved using two $\beta$-barium-borate (BBO) crystals.
We also measured the coincidence counts of generated photon pairs 
in a non-collinear configuration.
Calculations based on the experimental results 
suggest that it should be possible to 
realize a bandwidth of approximately 215 nm (100 THz)
by using a set of four BBO crystals.

We also discuss the effect of air gaps between multiple NLCs on the Hong-Ou-Mandel (HOM) interference \cite{hong1987measurement} between the daughter photons generated by our scheme. Our theoretical analysis suggests that the difference in the refractive indices of the air for the pump light and the daughter photons indeed modify the shape of the HOM interference, but the change is negligibly small for the current experimental condition. The analysis also suggests that the effect of group velocity mismatch between the pump light and the daughter photon is also negligible and it is not necessary to stabilize the distance between crystals with interferometric accuracy when a pump light with appropriate coherence length is used.

The remainder of the paper is organized as follows. 
Section 2 describes the proposed scheme.
Section 3 describes the experimental setup used 
to demonstrate the scheme by using two bulk NLCs.
Section 4 presents the experimental results 
and discussions about the effect of the air gap between NLCs.
The final section summarizes the findings of this study 
and gives the conclusions.

\section{Proposed scheme}

For bulk NLCs,
the spontaneous parametric fluorescence spectrum is determined by 
the phase-matching conditions for parametric down conversion, 
which is a second-order nonlinear process.
A pump photon with a frequency $\omega_p$ and a momentum $k_p$
is down converted to a pair of photons, which are signal and idler photons
with frequencies $\omega_s$ and $\omega_i$ 
and momenta $k_s$ and $k_i$, respectively.
Due to energy conservation, $\omega_p=\omega_s + \omega_i$ must be satisfied
so that the momentum mismatch $\Delta k=k_p - k_s - k_i$ must be zero
to satisfy the critical phase-matching condition \cite{boyd2003nonlinear}.
The frequency spectrum $F(\omega)$ for spontaneous parametric fluorescence,
which consists of pairs of signal $F(\omega_s)$ and idler $F(\omega_i)$ photons,
can be assumed to be symmetric relative to the center frequency $\frac{\omega_p}{2}$.
The spectrum $F(\omega)$, 
which is $F(\omega_s)=F(\omega_i)=F(\frac{\omega_p}{2} \pm \Omega)=F(\Omega)$,
can then be described as
\begin{equation}
F(\Omega) \propto \left| \int^{L}_{0}dz \exp\left[i \Delta k(\Omega) z\right] \right|^2
= L^2 \textrm{sinc}^2 \left[\frac{\Delta k(\Omega)L}{2} \right],
\end{equation}
where $z$ is the position in the NLC along the pump beam axis.
 and $L$ is the length of the NLC along the $z$ axis.
The momentum mismatch $\Delta k$ in the NLC depends on the frequencies of the
generated signal and idler photons.
Thus, the frequency spectrum $F(\omega)$ has a finite bandwidth that is
determined by the frequency-dependent momentum mismatch $\Delta k(\Omega)$.

\begin{figure}[ht]
\centering
\includegraphics[width=11cm]{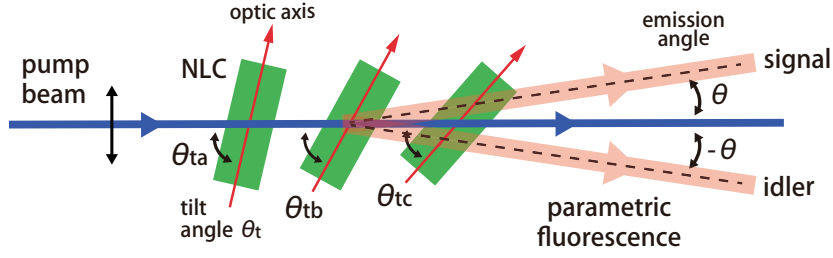}
\caption{Schematic of proposed method to generate spontaneous parametric fluorescence
by using multiple bulk NLCs.
Multiple (three in figure) bulk NLCs are aligned in parallel; 
the pump beam passes through all the NLCs.
The optic axes of the NLCs have different tilt angles $\theta_t$
($\theta_{ta}, \theta_{tb}, \theta_{tc}$) 
to control the phase-matching condition for each NLC.
Generated parametric fluorescence is collected
at the emission angle $\pm\theta$ as signal and idler photons.}
\label{schematic}
\end{figure}

We now propose our scheme that employs multiple NLCs
to generate broadband spontaneous parametric fluorescence.
As an example, Fig. \ref{schematic} schematically depicts the proposed method for three NLCs.
Multiple bulk NLCs are aligned in parallel  along the pump beam.
The phase-matching condition, which depends on the momentum mismatch $\Delta k$,
for each NLC can be controlled 
by controlling the tilt angle $\theta_t$ of the optic axis of the NLC.
The tilt angle between the optic axis and the pump beam 
is varied in the horizontal plane (i.e., parallel to the page) and
the pump beam is horizontally polarized in Fig. \ref{schematic}. 
The tilt angles of the three NLCs are set to be different from each other 
($\theta_{ta}, \theta_{tb}$, and $\theta_{tc}$ in Fig. \ref{schematic}).
Consequently, the parametric fluorescence spectra generated by the NLCs
will differ due to the phase-matching conditions.
The parametric fluorescence generated from all the NLCs can be collected
at an emission angle $\theta(-\theta)$ as signal (idler) photons of correlated photon pairs
in a non-collinear configuration.
The parametric fluorescence spectrum also depends on the emission angle $\theta$,
which determines the momentum direction of generated photons.
Thus, the signal photon spectra generated from the NLCs 
and detected at an emission angle $\theta=\theta_d$, can be expressed by
\begin{equation}
F(\Omega, \theta_d) = \sum_{i=a,b,c} F_i(\Omega, \theta_d, \theta_{ti})
\propto \sum_{i=a,b,c} L^2 \textrm{sinc}^2 
\left[\frac{\Delta k_i(\Omega, \theta_d, \theta_{ti})L}{2} \right],
\end{equation}
where $F_i, \theta_{ti}$, and $k_i$ are the signal photon spectrum, 
the NLC tilt angle, and the momentum mismatch
of the $i$th NLC ($i=a, b, c$), respectively.
By suitably controlling the phase-matching conditions of the NLCs,
the spontaneous parametric fluorescence bandwidth could be broadened 
due to the superposition of spontaneous parametric fluorescence 
generated from the multiple bulk NLCs.

\begin{figure}[ht]
\centering
\includegraphics[width=11.5 cm]{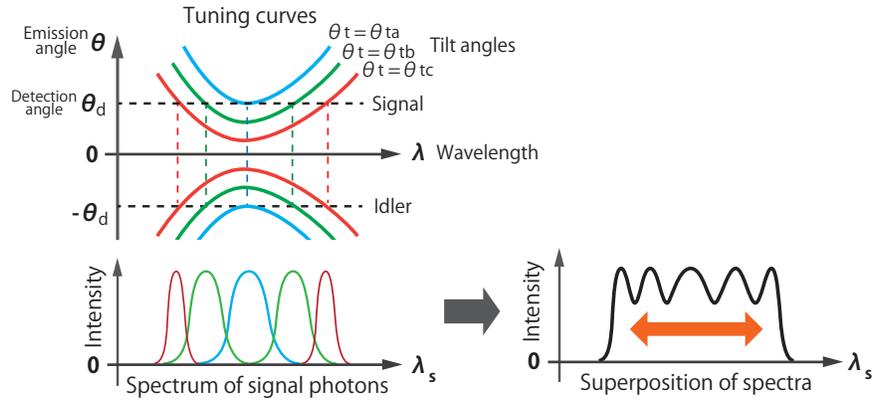}
\caption{Illustration of proposed scheme for 
generating spontaneous parametric fluorescence
using three bulk NLCs.
Upper left figure shows tuning curves of three NLCs with
tilt angles $\theta_t$ of $\theta_{ta}, \theta_{tb}$, and $\theta_{tc}$. 
Lower left figure shows corresponding 
parametric fluorescence spectra from these three NLCs
for a detection angle $\theta=\pm\theta_d$.
Lower right figure shows broadened spectrum
as a superposition of these spectra.}
\label{spectrum schematic}
\end{figure}

Figure \ref{spectrum schematic} schematically depicts the proposed scheme with three NLCs
in the view of the frequency spectrum. 
The upper left figure of Fig. \ref{spectrum schematic} shows 
the relation between the emission angle $\theta$ and the wavelength $\lambda$ 
of the generated parametric fluorescence 
as tuning curves that indicate the phase-matching conditions. 
The three different tilt angles $\theta_t$ of the
three NLCs ($\theta_{ta}, \theta_{tb}, \theta_{tc}$)
lead to three different tuning curves.
When the generated signal and idler photons are collected 
at emission angles of $\theta=\theta_d$ and $-\theta_d$, respectively,
the parametric fluorescence spectra generated by the NLCs
have different peaks and bandwidths, as shown in the lower left figure of Fig. \ref{spectrum schematic}.
The broadband spontaneous parametric fluorescence can be obtained 
at the detection angle $\theta_d$ as a superposition of these spectra,
as shown in the lower right figure of Fig. \ref{spectrum schematic}. 
Following the proposed method,
the spectral bandwidth can be broadened 
by increasing the number of crystals.
The expected broadening of the bandwidth 
with increasing number of crystals
will be discussed later in Sec. 4.1 
based on experimentally measured spectra.

\section{Experimental setup}

\begin{figure}[hbt]
\centering
\includegraphics[width=12 cm]{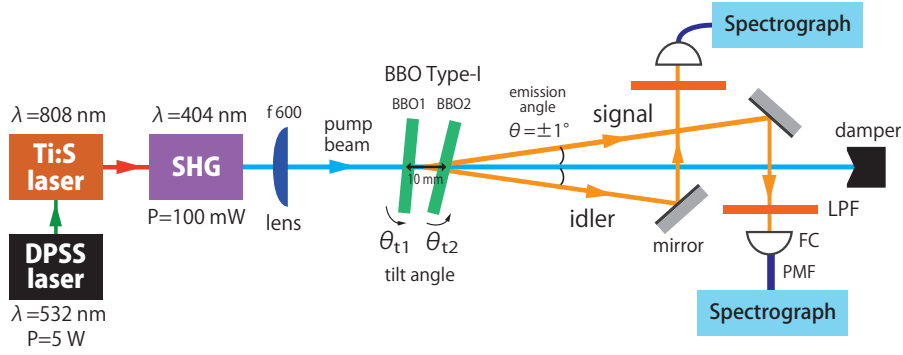}
\caption{Experimental setup to measure 
spontaneous parametric fluorescence spectra generated from one or two BBO crystals (BBO1 and BBO2). 
$\theta_{t1}$ and $\theta_{t2}$ are the tilt angles of 
the optic axes of BBO1 and BBO2 crystals relative to the cut angle, respectively.
The two BBO crystals are aligned in parallel along  the pump beam with a separation of 10 mm.
Generated signal and idler photons are collected at an emission angle of $\pm$1$^{\circ}$
by fiber couplers and are transferred to the spectrograph to obtain the spectra.
DPSS laser: diode-pumped solid-state laser, SHG: second-harmonic generation system,
LPF: long pass filter, FC: fiber coupler, PMF: polarization-maintaining fiber.}
\label{experimental setup}
\end{figure}

Figure \ref{experimental setup} shows the experimental setup 
used to demonstrate the proposed scheme using two bulk NLCs.
The pump laser system consists of a single-frequency CW Ti:sapphire laser
(MBR-110, Coherent) pumped by a diode-pumped solid state (DPSS) laser 
(Verdi G-10, Coherent) and a resonant frequency doubling unit (MBD-200, Coherent).
The output of the Ti:sapphire laser (wavelength: 808 nm; 
linewidth: approximately 100 kHz) is frequency doubled 
by second-harmonic generation (SHG) and it is used as the  pump beam.
The reason why we used a pump laser with narrow linewidth 
(coherence length longer than 1 km) will be discussed later in Sec. 4.2.
The pump beam (wavelength: 404 nm; 
power: 100 mW) is weakly focused by a lens (focal length: 600 mm);
this focusing has a confocal parameter (i.e., twice the Rayleigh length) of approximately 230 mm.
Two 2-mm-thick bulk BBO crystals (BBO1 and BBO2)
are aligned in parallel along  the pump beam with a separation of 10 mm around the focus.
The optic axis of BBO1 (BBO2) crystal has a tilt angle 
of $\theta_{t1}$ ($\theta_{t2}$) 
relative to the cut angle (28.9$^{\circ}$) of the BBO crystal. 
The signal and idler photons generated by type-I phase-matching parametric down conversion  
are collected by two fiber couplers (FCs) 
with an emission angle $\theta$ of 1$^{\circ}$ relative to the pump beam.
Long pass filters (LPFs) are placed in front of the FCs to filter the pump beam.
Collected photons are transferred 
through polarization-maintaining fibers (PMFs)
to the 300-mm spectrograph with a 300-grooves/mm grating blazed at 750 nm
(SP-2358, Princeton Instruments) 
and detected by a charge coupled device (Pixis:100BRX, Princeton Instruments)
to measure the frequency spectrum.
The phase-matching conditions of the two BBO crystals
are controlled by rotating the BBO crystals
to vary the tilt angles $\theta_{t1}$ and $\theta_{t2}$.

\section{Results and discussion}

\subsection{Observation of broadband spontaneous parametric fluorescence}

\begin{figure}[htb]
\label{experimental_results}
\centering
\includegraphics[width=13 cm]{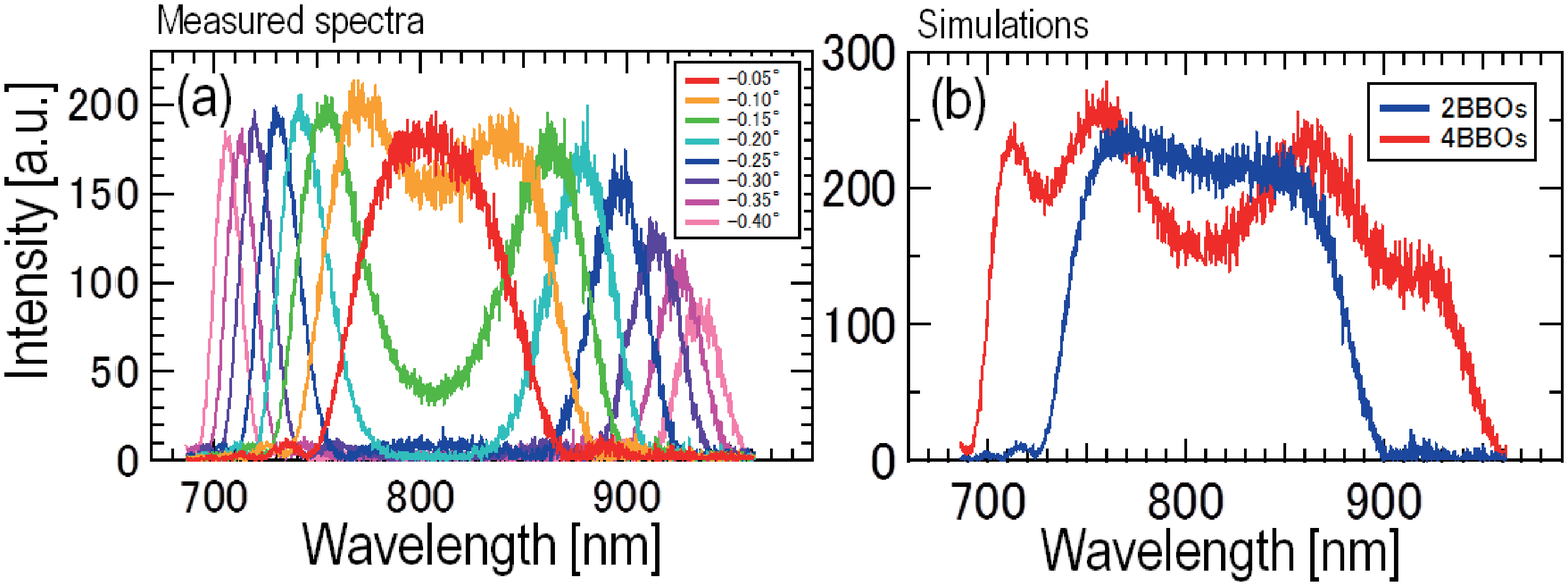}
\caption{(a) Measured spontaneous parametric fluorescence spectra
generated from one BBO crystal (BBO2) 
with a tilt angle $\theta_{t2}$ varied between $-0.05^{\circ}$ and $-0.40^{\circ}$.
(b) Calculated spectra as a superposition of measured spectra
of two BBO crystals with $\theta_{t2}$ of 
$-0.05^{\circ}$ and $-0.15^{\circ}$ (blue line)
and four BBO crystals with $\theta_{t2}$ of 
$-0.10$, $-0.20$, $-0.30$, and $-0.40^{\circ}$ (red line).
}
\end{figure}

To experimentally demonstrate our scheme, 
we measured spontaneous parametric fluorescence spectra generated 
from one and two NLCs in the experimental setup 
shown in Fig. \ref{experimental setup}.
We first measured spontaneous parametric fluorescence spectra 
generated from only one BBO crystal (BBO2)
in a non-collinear configuration for various phase-matching conditions,
as shown in Fig. \ref{experimental_results}(a).
The tilt angle $\theta_{t2}$ was varied from $-0.05^{\circ}$ to $-0.40^{\circ}$
in 0.05$^{\circ}$ steps.
As a result, the spectrum could be controlled 
by varying the phase-matching conditions,
as shown in Fig. \ref{spectrum schematic}.
The transmission efficiency of the spectrograph was calibrated in these spectra;
however, the intensity reduction in the long wavelength region may be
due to a reduction in the fiber coupling efficiency. 
When the tilt angle $\theta_{t2}$ was set to $-0.05^{\circ}$,
the spectrum was degenerate at the center wavelength of 808 nm 
and the bandwidth (FWHM) was typically 75 nm (34 THz).
Thus, a broadband spectrum is expected
when spectra with different tilt angles are superimposed.
As an example, Fig. \ref{experimental_results}(b) 
shows calculated spectra as a superposition of these measured spectra.
The bandwidth of the spectra obtained by summing two measured spectra
obtained for tilt angles $\theta_{t2}$ of $-0.05^{\circ}$ and $-0.15^{\circ}$ 
is approximately 130 nm (60 THz), as plotted by the blue line 
in Fig. \ref{experimental_results}(b).
A bandwidth of approximately 215 nm (100 THz),
which is almost three times that of one degenerate spectrum,
can be expected from the sum of four measured spectra 
with $\theta_{t2}$ of $-0.10^{\circ}$, $-0.20^{\circ}$, 
$-0.30^{\circ}$, and $-0.40^{\circ}$,
as plotted by the red line in Fig. \ref{experimental_results}(b).

\begin{figure}[hbt]
\centering
\includegraphics[width=10cm]{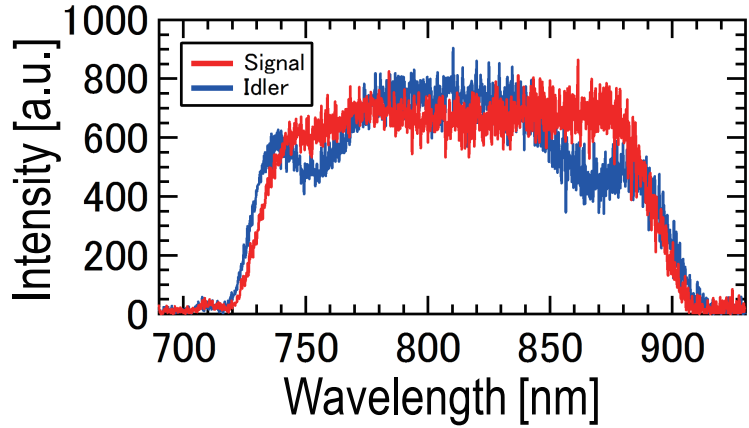}
\caption{Measured spectra of signal (red line) and idler (blue line) photons
of spontaneous parametric fluorescence generated from two BBO crystals.
Spectra of signal and idler photons had measured bandwidths of 
approximately 160 nm (73 THz).
Tilt angles $\theta_{t1}$ and $\theta_{t2}$ 
were set to $-0.10^{\circ}$ and $-0.20^{\circ}$, respectively.
}
\label{exp_data_fig5}
\end{figure}

We then measured spectra of both the signal and idler photons of the
spontaneous parametric fluorescence generated from both BBO crystals
as shown in Fig. \ref{exp_data_fig5}.
A typical bandwidth of 160 nm (73 THz) was obtained 
with tilt angles $\theta_{t1}$ and $\theta_{t2}$ 
of $-0.10^{\circ}$ and $-0.20^{\circ}$, respectively
as shown in Fig. \ref{exp_data_fig5}.
A typical single photon count rate was 1.0 $\times$ 10$^5$ counts per second
when the pump beam power was 10 mW.
The measured bandwidth was over two times greater than that 
obtained using only one BBO crystal.
This broad bandwidth in the frequency domain corresponds to 
a Fourier-transform-limited temporal width of approximately 6 fs 
for correlated photon pairs in the time domain.
The intensity ratio of parametric fluorescence collected
from two BBO crystals was sensitive to the fiber coupling of the fiber couplers, 
which was optimized to maximize the overlap of these two spectra.
The slight difference between the spectrum of 
signal (red line in Fig. \ref{exp_data_fig5}) and 
idler (blue line in Fig. \ref{exp_data_fig5}) photons
may be caused by the non-perfect spatial mode matching
of the two detection modes at the crystal.

Then we measured the coincidence counts 
between the signal photons and the idler photons 
for the whole spectrum, i.e. 
just inserting a low frequency filter (SCF-50S-42L, SIGMA KOKI) 
to cut the pump beam.
With the detection event of the signal photon as a trigger,
the coincidence count rate of idler photons was typically 2\%
that of the single photon count rate.
We think this relatively low fraction (2\%)
is due to technical difficulties 
when we align single mode fiber couplers.
We think the fraction can be improved up to approximately 20\%
if we can simultaneously monitor the spectra of daughter photons
and coincidence counting rate
\cite{baek2008spectral}.

\subsection{Effect of the air gap between the nonlinear crystals
in the proposed scheme}

In our proposed scheme, 
there should be gaps between NLCs to control the tilt angles of NLCs, which may affect the phase relation 
between generated photons and pump photons due to following two causes.

The first cause is the group velocity mismatch 
between the pump beam and signal (idler) photons.
The group velocity mismatch can be evaluated by the dispersion coefficient
$D=u_o^{-1}-u_e^{-1}$, where $u_{o,e}$ are the group velocities 
for ordinary and extraordinary rays in a medium \cite{Giuseppe2002airgap}.
Then the spatial separation $\Delta l$ of the pump and signal (idler) photons
after passing through the medium with a length $d$ 
can be given by $\Delta l=cdD$.
If the coherence length of the pump beam is shorter than this separation,
this separation causes a loss of the coherence between photons
generated from separated multiple NLCs and 
thus only an incoherent mixture of photons can be obtained \cite{{Kim2000},{Rangarajan2009}}.

In our experimental setup,
two BBO crystals with a length of 2 mm is separated 
by the air gap with a distance of 10 mm.
In this case, the separation $\Delta l$ in the BBO crystal is approximately 0.1 mm 
and that in the air gap is less than 1 $\mu$m.
In our experimental setup shown in Sec. 3, the CW pump laser with a narrow linewidth ($\sim$ 100 kHz) is used. The 
coherence length of the laser is an order of 1 km and is much longer than these separations ( 0.1 mm and 1 $\mu$m).
In this condition, the effect of group velocity mismatch is negligible and it is not necessary to stabilize the distance between crystals with interferometric accuracy.
Note that similar condition has been widely used for the polarization-entanglement sources \cite{{Kwiat1999},{Kawabe2007},{Fujiwara2011}}. This effect of group velocity mismatch will become critical when one wish to use short pulse lasers with small coherent length for pumping \cite{{Kim2000},{Rangarajan2009}}.

The second cause is the momentum mismatch  between the pump, signal and idler photons. This effect on the polarization entangled photon pair source has been discussed 
by Atat\"{u}re et al. \cite{Atature2001cascaded}
and Giuseppe et al. \cite{Giuseppe2002airgap}.
For the case of frequency entangled photon pairs generated from the proposed scheme in Sec. 2,  the state of photon pairs generated from two NLCs is written by 
\begin{equation}
|\Psi>= \int d\Omega F_a(\Omega)|+\Omega>_s|-\Omega>_i
+\int d\Omega \exp\left[i \phi_d(\Omega)\right]
F_b(\Omega)|+\Omega>_s|-\Omega>_i,
\label{photonstate}
\end{equation}
where $F_i(+\Omega)=F_i(-\Omega)$ $(i=a,b)$ is a symmetric frequency spectrum
of photons generated from $i$th NLC as explained in Sec. 2 and
$\phi_d(\Omega)$ is a phase term.
This phase term can be given by $\phi_d(\Omega)=\Delta k(\Omega) d$,
where $\Delta k(\Omega)$ is a momentum mismatch 
between the pump beam, signal and idler photons in the air gap.
This momentum mismatch can be written by
$
\Delta k(\Omega)=k_p-k_s-k_i
=n(\omega_p)\omega_p/c
-n(\frac{\omega_p}{2}+\Omega)(\frac{\omega_p}{2}+\Omega)/c
-n(\frac{\omega_p}{2}-\Omega)(\frac{\omega_p}{2}-\Omega)/c,
$
where $n(\omega)$ is a refractive index of the air at the frequency $\omega$.

To investigate the effect of this phase term,
Here we consider a HOM interferometer 
\cite{hong1987measurement},
with a pair of photons described in Eq.(\ref{photonstate}).
The coincidence rate $P_c(\tau)$ 
between two separated output ports A and B of the HOM interferometer
with the slow single photon detectors can be written by 
\begin{eqnarray}
P_c(\tau) &\propto& 
\int d\omega_A d\omega_B
<\Psi|a_A^{\dagger}(\omega_A)a_B^{\dagger}(\omega_B)
a_A(\omega_A)a_B(\omega_B)|\Psi>\\
&=& \int d\Omega_A d\Omega_B
||<0|a_s(\Omega_A)a_i(\Omega_B)e^{i\Omega_B \tau}
-a_s(\Omega_B)a_i(\Omega_A)e^{i\Omega_A \tau}|\Psi>||^2\\
&=& \int d\Omega 
|F_a(\Omega)+\exp\left[i \phi_d(\Omega) \right] F_b(\Omega)|^2
\left( 1-\cos (2\Omega \tau) \right),
\label{HOMdip}
\end{eqnarray}
where $\tau$ is a relative time delay between signal and idler photons,
$a_j(\omega_j)(j=A,B)$ is an annihilation operator 
for the mode coupled to the output port $j=A,B$ at the frequency $\omega_j$
and $a_j(\Omega_k)(j=s,i$ and $k=A,B)$ is an annihilation operator 
for the mode of signal ($j=s$) and idler ($j=i$) photons 
at the frequency $\Omega_k$ \cite{Steinberg1992}.

\begin{figure}[htb]
\centering
\includegraphics[width=12.5 cm]{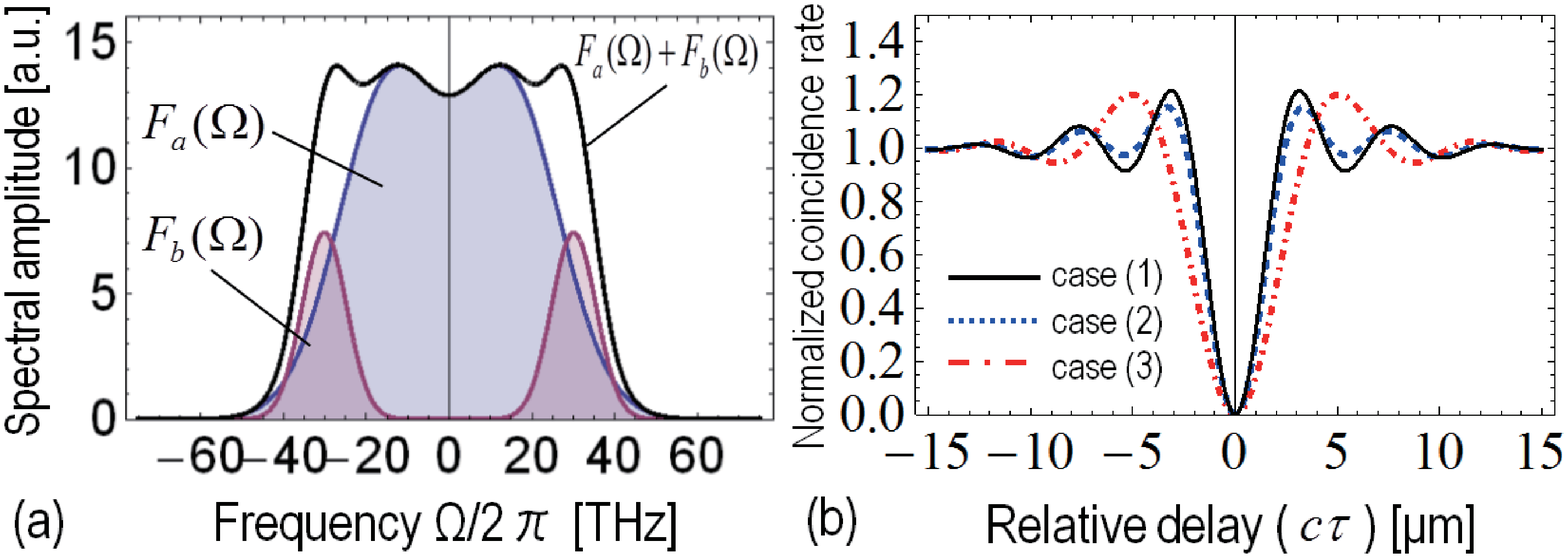}
\caption{(a) Frequency spectra $F_i(\Omega)$ of photons generated from
$i$th NLC ($i=a$ (blue line), $b$ (red line)) 
and a sum of two spectra (black line)
with the bandwidth of approximately 70 THz.
(b) Coincidence rate $P_c(\tau)$ 
in a HOM two-photon interferometer
with the case (1) $\phi_d = 0$ (solid black line), 
case (2) $\phi_d \sim 0.37 \pi$ (dashed blue line) 
and case (3) $\phi_d = \pi$ (dashed-dotted red line).
The width of the HOM dip with the case (1) is
approximately 3 $\mu$m}.
\label{exp_data_fig6}
\end{figure}

Now, let us consider HOM interference signals
with the state of photons that have frequency spectra shown in Fig. \ref{exp_data_fig6}(a), which is a model spectrum based on the experimental data shown in Fig. 4.
The spectrum $F_{i}(\Omega)(i=a,b)$ of photons generated from $i$th NLC 
is assumed to be different from each other (cf. Fig. 2).
Then a sum of two spectra $F_a(\Omega)+F_b(\Omega)$ 
has broadened bandwidth ($\Omega/2\pi \sim 70$ THz in this case).

Figure \ref{exp_data_fig6}(b) shows
calculated two-photon interference signals of the HOM interferometer
for the following three cases: (1) An ideal case with the 0 gap distance ($\phi_d$ = 0);
(2) The case of our experimental condition, with
the pump laser wavelength $\lambda_p$ of 404 nm and 
the air gap distance $d$ of 10 mm
(The average of $\phi_d$ over the fluorescence spectrum is approximately 0.37 $\pi$);
(3) The worst case ($\phi_d = \pi$), which corresponds to the air gap distance approximately 27 mm for $\lambda_p$ of 404 nm.
Note that the phase term $\phi_d$
is a function of the gap distance $d$ between the crystals and 
refractive indices $n$ of the pump and signal (idler) photons. 
The results of the calculations for these three cases 
are shown in Fig. \ref{exp_data_fig6}(b). The solid black line, the dashed blue line, and the dashed-dotted red line corresponds to case (1), (2), and (3) respectively.
In case (1),
the HOM interference curve is simply the inverse Fourier transform
of the absolute value squared of the sum of the two spectra
($F_a(\Omega)+F_b(\Omega)$),
which is shown as the black line in  Fig. \ref{exp_data_fig6}(a).
The width of this HOM dip ($c\tau \sim 3$ $\mu$m) corresponds to the time correlation of the signal and the idler photon ($\tau \sim$ 9 fs). 
Note that the difference between the dashed blue line (case (2)) and 
the solid black line (case (1)) is negligibly small. 
In the worst case (case (3)), the width of the HOM dip is
almost the same with the HOM dip given solely by the first crystal ($F_a(\Omega)$).
Note that the difference between the curves becomes less significant when the overlap of $F_a(\Omega)$ and $F_b(\Omega)$ become smaller.
Note also that this effect caused by momentum mismatch changes has a sinusoidal dependence on the gap distance $d$, with a period of 54 mm in our experimental condition. Thus, it is not necessary to stabilize the distance between crystals with interferometric accuracy for this effect, too.

In summary in this subsection, 
we have considered the effect of group velocity mismatch and the momentum mismatch between the pump light and daughter photons on the shape of HOM interference curve. The effect of group velocity mismatch between the pump light and the daughter photon is negligible in the current experimental condition where a CW pump laser with long coherent length is used. The difference in the refractive indices of the air for the pump light and the daughter photons causes the momentum mismatch between the pump light and daughter photons, and it may modify the shape of the HOM interference, but the change is negligibly small for the current experimental condition. 
The calculated HOM interference curve using a model spectrum considering the experimental data is almost the same with the one calculated for an ideal frequency entangled photon pair. 
For both of the effects, 
it is not necessary to stabilize the distance between crystals with interferometric accuracy.

\section{Conclusion}

We proposed a novel method for 
generating broadband spontaneous parametric fluorescence
by using a set of multiple NLCs.
We demonstrated the proposed scheme using two conventional NLCs. 
The superposition of spontaneous parametric fluorescence spectra
generated from two BBO crystals gave a
typical bandwidth of 160 nm (73 THz) 
at the degenerate wavelength of 808 nm. 
In contrast, the bandwidth with only one BBO crystal was typically 75 nm (34 THz). 
We also measured coincidence counts of correlated photon pairs
generated in a non-collinear configuration.
Using this simple scheme, 
a bandwidth could be broadened by increasing the number of NLCs
without employing any special devices,
provided the phase-matching conditions of the NLCs are satisfied.
Based on a calculation using measured spectra, a bandwidth of
approximately 215 nm (100 THz) is expected
by using a set of four BBO crystals. 
Furthermore, the pump laser frequency can be tuned 
within the tuning range of the NLCs in the proposed method. 
This tunability is important for applications 
such as two-photon absorption experiments.

We have also discussed the effect of air gaps between multiple NLCs on the HOM interference between the daughter photons generated by our scheme. Our theoretical analysis suggested that the difference in the refractive indices of the air for the pump light and the daughter photons may modify the shape of the HOM interference, but the change is negligibly small for the current experimental condition. The analysis also suggested that the effect of group velocity mismatch between the pump light and the daughter photon is also negligible and it is not necessary to stabilize the distance between crystals with interferometric accuracy when a pump light with appropriate coherence length is used.
The narrow HOM dips of broadband spontaneous parametric fluorescence generated in our scheme are promising for applications such as a quantum optical coherence tomography.

For the generation of broadband parametric fluorescence, 
higher efficiency may be expected 
when periodically polled materials are used 
due to the higher nonlinearities. 
On the other hand, our scheme has advances 
on the tunability of both pump and signal/idler wavelengths 
by simply adjusting the angles of the crystals. 
It may be also possible to prepare crystals 
cut at different angles and put together in contact 
to reduce the dispersion effect 
when tunability is not necessary. 
The limit of the bandwidth is, in principle, 
given by the phase matching condition 
and transparent window of the nonlinear crystal. 
However, the increase in number of crystals 
to achieve very broad spectrum 
may suffer from technical difficulties 
like fine adjustment of each crystals.

Our scheme could also be useful for conventional nonlinear optics
especially with the process of the second harmonic generation,
which is an inverse process of the parametric down-conversion.
The second harmonic generation 
from broadband anti-correlated frequencies 
have recently been demonstrated by classical laser pulses
for optical coherence tomography
\cite{{Peer2007sfg},{Kaltenbaek2008coct}}.

\section*{Acknowledgments}

This work was supported in part by Grant-in-Aids from the 
JST-CREST Project, JSPS, the Quantum Cybernetics Project, 
the FIRST Program of JSPS, 
the Project for Developing Innovation Systems of MEXT, 
the GCOE program, 
and the Research Foundation for Opto-Science and Technology.

\end{document}